\documentclass{ws-procs9x6}

\begin{document}

\newcommand{\vecr}{\mbox{\boldmath $r$}}
\newcommand{\FSA}{8cm}
\newcommand{\FSB}{16cm}
\newcommand{\RMEY}{\left\langle 
  \left( {\textstyle \frac{1}{2}} l  \right) j \right. 
    \left\|  {\bf Y}_L \left\| 
  \left( {\textstyle \frac{1}{2}} l' \right) j' \right. 
\right\rangle}
\newcommand{\UVLM}{
\left[  U_{nlj} (\vecr) V_{n'l'j'} (\vecr) \right]_M^{(L)}  }
\newcommand{\UVLMS}{
\left[  U_{nlj}^{*} (\vecr) V_{n'l'j'}^{*} (\vecr) \right]_M^{(L)}  }
\newcommand{\FPL}{\frac{\sqrt{4\pi}}{\hat{L}}} 
\newcommand{\textfrac}[2]{{\textstyle \frac{#1}{#2}}}
\newcommand{\sigmab}{\bar{\sigma}}
\newcommand{\mcU}{\mathcal{U}}
\newcommand{\mcV}{\mathcal{V}}
\newcommand{\betwo}{$B(E2, 0^{+}_1 \rightarrow 2^{+}_1)$ }
\newcommand{\beq}{$B(Q^{n}2;0^+_1 \rightarrow 2^+_1)$}
\newcommand{\twop}{$2^{+}$ }
\newcommand{\Mg}{$^{32}$Mg}
\newcommand{\Ne}{$^{30}$Ne}


\title{A self-consistent QRPA study of 
quadrupole collectivity around $^{32}$Mg}

\author{Masayuki Yamagami}
\address{Department of Physics, Graduate School of Science,\\
Kyoto University, Kyoto 606-8502, Japan\\
E-mail:yamagami@ruby.scphys.kyoto-u.ac.jp}  

\author{Nguyen Van Giai}
\address{Institut de Physique Nucl\'eaire, IN$_{2}$P$_{3}$-CNRS,\\
91406 Orsay Cedex, France\\
E-mail:nguyen@ipno.in2p3.fr}  


\maketitle

\abstracts{On the basis of the Hartree-Fock-Bogoliubov (HFB) plus
quasiparticle random phase approximation method (QRPA) based on 
the Green's function approach with Skyrme force, 
we discuss the anomalous E2 properties of the first 2$^{+}$ states 
in neutron-rich nuclei \Mg\ and \Ne.
The B(E2) values and the excitation energies 
of the first 2$^{+}$ states are well described 
within HFB plus QRPA calculations with spherical symmetry. 
We conclude that pairing effects account largely for the anomalously large B(E2) 
values and the low excitation energies in \Mg\ and \Ne.
}


\section{Introduction}

The observed anomalous E2 properties in \Mg\ and \Ne, 
the large B(E2) values and the low excitation energies, 
are clear evidences of the vanishing of the N=20 shell closure.\cite{MI95,YN03}
Several theoretical studies have shown the importance of the neutron 2p-2h 
configurations across the N=20 shell gap to describe these anomalous 
properties (e.g.[3]).
The 2p-2h configurations imply deformation of these nuclei, however, 
it is under great debate whether \Mg\ is deformed or not.
The observed energy ratios $E(4^+_1)/E(2^+_1)$ is 2.6 
in \Mg\ [4,5]. 
This value is in between the rigid rotor limit 3.3 and 
the vibrational limit 2.0. 
The B(E2) value (in single-particle units) is 
15.0$\pm$2.5 in $^{32}$Mg, and
this value is smaller than in deformed Mg isotopes 
(21.0$\pm$5.8 in $^{24}$Mg, 19.2$\pm$3.8 in $^{34}$Mg [6]).
Moreover, in mean-field calculations,
irrespective to relativistic or non-relativistic, 
the calculated ground states in $^{32}$Mg 
have been found to be spherical (e.g.[7]). 
Generally speaking, the neutron 2p-2h configurations can originate
not only from deformation but also from neutron pairing correlations.
In \Mg\ these two effects may coexist and help to make the anomalous 
E2 properties. In the previous studies it is not clear 
which effect is more essential to describe the anomalous properties. 

We have performed HFB plus QRPA calculations with Skyrme force 
for the first 2$^{+}$ states in N=20 isotones.\cite{YG03}
The QRPA equations are solved in coordinate space by using the Green's function 
method.\cite{YG03,KS02} To emphasize the role of neutron pairing correlation, 
spherical symmetry is imposed. The residual interaction is 
consistently derived from the hamiltonian density 
of Skyrme force that has an explicit velocity dependence.
A detailed account of the method can be found in Ref.[8].
We obtained a good agreement not only qualitatively 
but also quantitatively with the experimental results.

\section{Ground state properties in N=20 isotones}

\begin{figure}[bt]
\centerline{\epsfxsize=4.5in\epsfbox{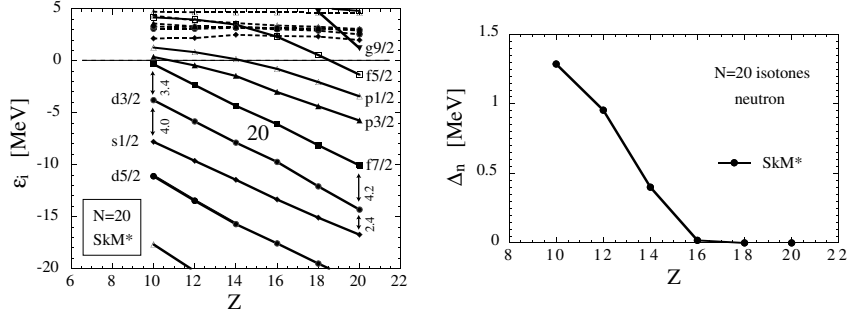}}   
\caption{HF neutron single-particle levels and the
neutron pairing gaps in N=20 isotones 
calculated with SkM*. Single-particle levels represented by 
solid lines correspond to bound and
resonance-like states, dashed lines to positive energy
discretized states. 
\label{FIG-GS}}
\end{figure}

The ground states are given by Skyrme-HFB calculations.
The Skyrme parameter SkM*
and the density-dependent pairing interaction,  
$
V_{pair} ( \vecr,\vecr') = V_{pair} 
[ 1 - \rho (\vecr) /\rho_c ]
\delta ( \vecr - \vecr'),
$
are adopted.  
$\rho_c=0.16$ fm$^{-3}$ is fixed. 
The strength $V_{pair}=-418$ MeV$\cdot$fm$^{-3}$
is determined so as  to reproduce the experimental neutron pairing gap in 
$^{30}$Ne. The quasiparticle cut-off energy is taken to be $E_{cut}=50$ MeV.
Fig.\ref{FIG-GS} shows the neutron single-particle levels 
in N=20 isotones calculated in HF.
The size of the N=20 shell gaps change slowly, 
because $2d_{3/2}$ and $1f_{7/2}$ orbits 
have high centrifugal barriers. On the other hand, 
the calculated neutron pairing gaps change considerably from 1.26 MeV 
in $^{30}$Ne to zero in $^{38}$Ar (Fig.\ref{FIG-GS}). 
The mechanism can be understood by the change of the level density 
in the $fp$ shell.
As close to the neutron drip-line, the single-particle energy (SPE) of
the high-{\it l} orbit $1f_{7/2}$ change almost linearly 
while the changes of $2p_{3/2}$ and $2p_{1/2}$ SPEs
become very slow. Moreover, the spin-orbit splitting
of $2p_{3/2}$ and $2p_{1/2}$ states becomes smaller.
Because of these different {\it l}-dependences of the SPEs,
the level density in the $fp$ shell becomes higher in \Mg\ and \Ne.
Within HFB calculations with spherical symmetry, the N=20 shell 
gap is naturally broken by neutron pairing correlations.

\begin{figure}[tb]
\centerline{\epsfxsize=4.2in\epsfbox{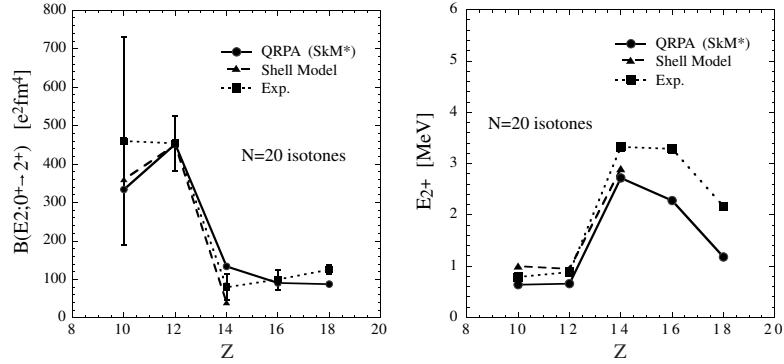}}   
\caption{
The \betwo transition probabilities and excitation energies 
of the first 2$^{+}$ states in N=20 isotones calculated in 
QRPA with SkM*. For comparison the available experimental data [1,2]
and the results of shell model [3] 
are shown.
\label{FIG-E2}}
\end{figure}

\section{E2 properties in N=20 isotones}

\begin{figure}[bt]
\centerline{\epsfxsize=4.2in\epsfbox{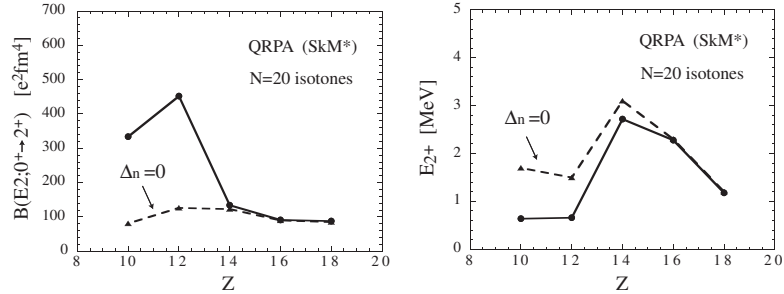}}   
\caption{The \betwo values and the excitation energies of the first 
  \twop states in N=20 isotones calculated with/without neutron pairing 
  correlations. Proton pairing is included in both cases.
\label{FIG-E2V0}}
\end{figure}

We have calculated the first \twop states in N=20 isotones 
in HFB plus QRPA calculations with spherical symmetry. 
Our aim is to investigate whether 
these \twop states can be described as vibrational states
built on the spherical ground states.
In Fig.\ref{FIG-E2} our QRPA results
are compared with the available experimental data\cite{MI95,YN03}
and the results of shell model.\cite{UO99}
The QRPA calculations have been done with SkM* and the fixed pairing strength.
The general properties of the first \twop states in N=20 isotones, 
especially large quadrupole collectivity 
in \Mg\ and \Ne, are well reproduced.
\begin{figure}[tb]
\centerline{\epsfxsize=4.2in\epsfbox{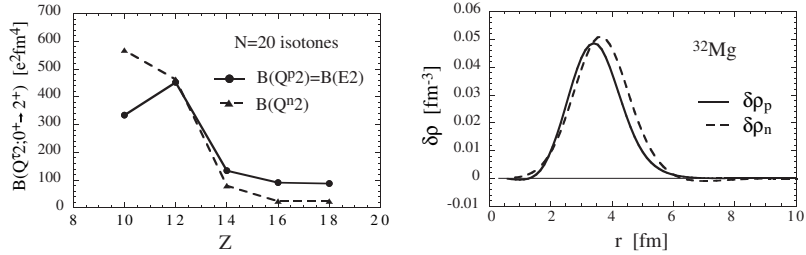}}   
\caption{The proton and neutron transition probabilities
$B(Q^{\tau}2;0^+ \rightarrow 2^+)$ in N=20 isotones, and
the transition densities in \Mg\ calculated by QRPA with SkM*.
\label{FIG-TRAN}}
\end{figure}
Without neutron pairing correlations, 
we cannot explain the anomalous E2 properties
(Fig.\ref{FIG-E2V0}).
Under these considerations, we can conclude that the large quadrupole 
collectivity in \Mg\ and \Ne\ appears thanks to 
the neutron pairing correlations. 
\begin{figure}[bt]
\centerline{\epsfxsize=1.95in\epsfbox{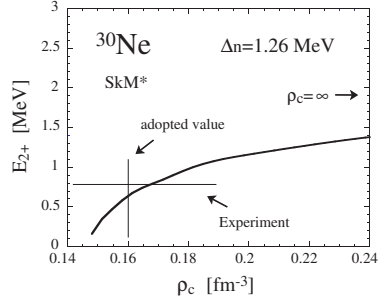}}   
\caption{The excitation energy of the first 2$^+$ state 
in \Ne\ as a function of $\rho_c$.
The pairing strength is determined so as to get the 
experimental pairing gap $\Delta_n =1.26$ MeV at each $\rho_c$. 
The limit $\rho_c = \infty$ corresponds to the volume-type pairing.
\label{FIG-RC}}
\end{figure}
To understand the mechanism that neutron pairing correlations help to make 
the large B(E2) values, we calculated the neutron 
transition probability B(Q$^{n}$2).
If neutron pairing correlations exist, the B(Q$^{n}$2) value can be large and 
the surface of neutron density becomes soft (Fig.\ref{FIG-TRAN}). 
In this situation, at the first stage, an electric external field makes 
the proton density of the spherical ground state vibrate in small amplitude.
This proton vibration makes 
neutrons vibrate by coherence between protons and neutrons. 
This neutron vibration can be very large, 
because neutron density is very soft thanks to the neutron 
pairing correlations. Finally, this large neutron vibration makes protons 
vibrate again by coherence between protons and neutrons, and 
this proton vibration becomes very large. 
The transition density in \Mg\
clearly exhibits this situation that 
the proton and neutron densities vibrate altogether coherently
(Fig.\ref{FIG-TRAN}). 
The peak position of the neutron transition density is 
slightly outside the nucleus due to the presence of neutron skin.
We expect that the nature of this vibrational state 
is sensitive to the surface properties. 
Fig.\ref{FIG-RC} shows the excitation energy of the first 2$^+$ state 
in \Ne\ as a function of $\rho_c$.
This state is very sensitive 
to $\rho_c$, and $\rho_c = 0.16$ fm$^{-3}$ gives $E_{2^+}=0.64$ MeV 
that is close to the experimental observation 0.791(26) MeV.\cite{YN03} 
The information of low-lying collective states in neutron-rich nuclei
may help to pin down the density dependence of pairing interactions. 

\section{Conclusions}

We have studied the first \twop states in N=20 isotones by 
the HFB plus QRPA with Skyrme force.
Because of the different behaviors of the neutron $1f$ and $2p$ 
orbits around zero energy, 
the neutron pairing correlations appear. 
This mechanism breaks the N=20 magicity in $^{32}$Mg and $^{30}$Ne.
Within QRPA calculations with spherical symmetry,  
the B(E2) values and the excitation energies of the first 
\twop states in N=20 isotones including $^{32}$Mg and $^{30}$Ne 
are well described. 
The existing experimental data are reproduced quantitatively.
The important role of the neutron pairing correlations is emphasized.
In the real $^{32}$Mg nucleus, both neutron pairing and deformation effects  
may coexist and help to make the large B(E2) value, but our calculation 
shows that neutron pairing correlations are essential.


%



%

%
%
%



\begin{thebibliography}{0}






\bibitem{MI95} T. Motobayashi, {\it et al.}, 
Phys. Lett. 346B (1995) 9.

\bibitem{YN03} Y. Yanagisawa, {\it et al.},
Phys. Lett. B 566 (2003) 84.

\bibitem{UO99} Y. Utsuno, T. Otsuka, T. Mizusaki, M. Honma, 
Phys. Rev. C60 (1999) 054315. 


\bibitem{YS01} K. Yoneda, {\it et al.}, 
Phys. Lett. 499B (2001) 233.

\bibitem{Gu02} D. Guillemaud-Mueller, 
Eur. Phys. J. A 13 (2002) 63.

\bibitem{IM01} H. Iwasaki, {\it et al.}. 
Phys. Lett. 522B (2001) 227.

\bibitem{RD99}
P. -G. Reinhard, {\it et al.}, 
Phys. Rev. C60 (1999) 014316.

\bibitem{YG03} M. Yamagami, Nguyen Van Giai,
preprint nucl-th/0307051. 

\bibitem{KS02} E. Khan, {\it et al.}, 
Phys. Rev. C66 (2002) 024309.



\end{thebibliography}
\end{document}